\documentstyle[prl,twocolumn, aps, eqsecnum,  epsf]{revtex}
\def\epsfig#1#2#3#4
         {
         \epsfysize=#2 \vbox{ \hglue#3 \epsfbox[#4]{#1} }
         }
\def\epsfigrot#1#2#3#4
         {
         \epsfxsize=#2
         \setbox\rotbox=\hbox to #2{\epsfbox[#4]{#1}}
         \vbox{\hglue#3 \rotl\rotbox}
         }
\newbox\rotbox

\begin{document}
%\draft
\title{On the Relevance of Disorder for Dirac Fermions with Imaginary Vector
Potentials}
\author{Andr\'e LeClair}
\address{Newman Laboratory, Cornell University, Ithaca, NY
14853.}
\date{\today}
\maketitle

\begin{abstract}

We consider the effects of disorder in a Dirac-like Hamiltonian.
In order to use conformal field theory techniques, we argue that one
should consider disorder in an imaginary vector potential.  
This affects significantly the signs of the lowest order $\beta$eta
functions.  We present evidence for the existence of two distinct
universality classes, depending on the relative strengths of the gauge
field verses impurity disorder strengths.  In one class
all disorder is driven irrelevant by the gauge field disorder.

\end{abstract}
\vskip 0.2cm
\pacs{PACS numbers: }
\narrowtext
%		DEFINITIONS FOR TEX
%
%%%%%%%%%%%%%%%%%%%%%%%%%%%%%%%%%%%%%%%%%%%%%%%%%%%%%%%%%%%%%%%
%
%
%\def\e{\'e}
%\def\ee{\`e}
%%%%%%%%%%%%%%%%%%%DEFINITIONS%%%%%%%%%%%%%%%%%%%%%%%%%%%%%%%%%
%
\def\beg{\begin{equation}}
\def\oti{{\otimes}}
\def\bra#1{{\langle #1 |  }}
\def\lb{ \left[ }
\def\rb{ \right]  }
\def\tilde{\widetilde}
\def\bar{\overline}
\def\hat{\widehat}
\def\*{\star}
\def\[{\left[}
\def\]{\right]}
\def\({\left(}		\def\BL{\Bigr(}
\def\){\right)}		\def\BR{\Bigr)}
	\def\BBL{\lb}
	\def\BBR{\rb}
%
%%%%%%%%%%%%%%%%%%%%%%%%%%%%%%%%%%%%%%%%%%%%%%%%%%%%%%%%%%%%%%%
%
\def\zb{{\bar{z} }}
\def\zbar{{\bar{z} }}
\def\frac#1#2{{#1 \over #2}}
\def\inv#1{{1 \over #1}}
\def\half{{1 \over 2}}
\def\d{\partial}
\def\der#1{{\partial \over \partial #1}}
\def\dd#1#2{{\partial #1 \over \partial #2}}
\def\vev#1{\langle #1 \rangle}
\def\ket#1{ | #1 \rangle}
\def\rvac{\hbox{$\vert 0\rangle$}}
\def\lvac{\hbox{$\langle 0 \vert $}}
\def\2pi{\hbox{$2\pi i$}}
\def\e#1{{\rm e}^{^{\textstyle #1}}}
\def\grad#1{\,\nabla\!_{{#1}}\,}
\def\dsl{\raise.15ex\hbox{/}\kern-.57em\partial}
\def\Dsl{\,\raise.15ex\hbox{/}\mkern-.13.5mu D}
%
%%%%%%%%%%%%%%%%%%%%GREEK LETTERS%%%%%%%%%%%%%%%%%%%%%%%%%%%%%%
%
\def\th{\theta}		\def\Th{\Theta}
\def\ga{\gamma}		\def\Ga{\Gamma}
\def\be{\beta}
\def\al{\alpha}
\def\ep{\epsilon}
\def\vep{\varepsilon}
\def\la{\lambda}	\def\La{\Lambda}
\def\de{\delta}		\def\De{\Delta}
\def\om{\omega}		\def\Om{\Omega}
\def\sig{\sigma}	\def\Sig{\Sigma}
\def\vphi{\varphi}
%
%%%%%%%%%%%%%%%%%%%CALIGRAPHIC LETTERS%%%%%%%%%%%%%%%%%%%%%%%%%
%
\def\CA{{\cal A}}	\def\CB{{\cal B}}	\def\CC{{\cal C}}
\def\CD{{\cal D}}	\def\CE{{\cal E}}	\def\CF{{\cal F}}
\def\CG{{\cal G}}	\def\CH{{\cal H}}	\def\CI{{\cal J}}
\def\CJ{{\cal J}}	\def\CK{{\cal K}}	\def\CL{{\cal L}}
\def\CM{{\cal M}}	\def\CN{{\cal N}}	\def\CO{{\cal O}}
\def\CP{{\cal P}}	\def\CQ{{\cal Q}}	\def\CR{{\cal R}}
\def\CS{{\cal S}}	\def\CT{{\cal T}}	\def\CU{{\cal U}}
\def\CV{{\cal V}}	\def\CW{{\cal W}}	\def\CX{{\cal X}}
\def\CY{{\cal Y}}	\def\CZ{{\cal Z}}

\def\rvac{\hbox{$\vert 0\rangle$}}
\def\lvac{\hbox{$\langle 0 \vert $}}
\def\comm#1#2{ \BBL\ #1\ ,\ #2 \BBR }
\def\2pi{\hbox{$2\pi i$}}
\def\e#1{{\rm e}^{^{\textstyle #1}}}
\def\grad#1{\,\nabla\!_{{#1}}\,}
\def\dsl{\raise.15ex\hbox{/}\kern-.57em\partial}
\def\Dsl{\,\raise.15ex\hbox{/}\mkern-.13.5mu D}
%
%%%%%%%%%%%%%%%%%%%%GREEK LETTERS%%%%%%%%%%%%%%%%%%%%%%%%%%%%%%
%
%%%%%%%%%%%%%%% MATH CHARACTERS %%%%%%%%%%%%%%%%%%%%%%%%%%%%
%
\font\numbers=cmss12
%\font\numbers=cmu10 scaled\magstep1
\font\upright=cmu10 scaled\magstep1
\def\stroke{\vrule height8pt width0.4pt depth-0.1pt}
\def\topfleck{\vrule height8pt width0.5pt depth-5.9pt}
\def\botfleck{\vrule height2pt width0.5pt depth0.1pt}
\def\Zmath{\vcenter{\hbox{\numbers\rlap{\rlap{Z}\kern
0.8pt\topfleck}\kern 2.2pt
                   \rlap Z\kern 6pt\botfleck\kern 1pt}}}
\def\Qmath{\vcenter{\hbox{\upright\rlap{\rlap{Q}\kern
                   3.8pt\stroke}\phantom{Q}}}}
\def\Nmath{\vcenter{\hbox{\upright\rlap{I}\kern 1.7pt N}}}
\def\Cmath{\vcenter{\hbox{\upright\rlap{\rlap{C}\kern
                   3.8pt\stroke}\phantom{C}}}}
\def\Rmath{\vcenter{\hbox{\upright\rlap{I}\kern 1.7pt R}}}
\def\Z{\ifmmode\Zmath\else$\Zmath$\fi}
\def\Q{\ifmmode\Qmath\else$\Qmath$\fi}
\def\N{\ifmmode\Nmath\else$\Nmath$\fi}
\def\C{\ifmmode\Cmath\else$\Cmath$\fi}
\def\R{\ifmmode\Rmath\else$\Rmath$\fi}
%%%%%%%%%%%%%%%%%%%%%%%%%%%%%%%%%%%%%%%%%%%%%%%%%%%%%%%%%%%%%%%%%
 %%%%%%%%%%%%%%%%%% END OF DEFINITIONS %%%%%%%%%%%%%%%%%%%%%%
 %%%%%%%%%%%%%%%%%%%%%%%%%%%%%%%%%%%%%%%%%%%%%%%%%

%\section{Introduction}

%Bla..Bla Bla\cite{BPZ}. 

%\begin{equation}
%\label{hamilo}
%H=H^{atom}+H^{field}+H^{int}
%\end{equation}
%with
%\begin{eqnarray}
%H^{field}&=&\frac{1}{2} \int_{-\infty}^\infty
%dx [(\partial_t \phi)^2+(\partial_x\phi)^2] ,
%\\ \nonumber
%H^{atom}&=&\frac{ \omega_0}{2} \sigma_3 ,
%{}~~~~~H^{int}=\frac{\beta}{2} \partial_t \phi (x_0) \left( \sigma_++
%\sigma_- \right),
%\label{hamili}
%\end{eqnarray}

%The hamiltonian (\ref{hamili}) 

%\vbox{
%\epsfysize=8cm
%\epsfxsize=8cm
%\epsffile{classlin.eps}
%\begin{figure}
%\caption[]{\label{fig1} Accuracy comparison for various
%form factor contributions for $g=1/5$.}
%\end{figure}}

%\begin{references}
%\bibitem{BPZ}  Belavin, Polyakov and Zamolodchickov.
%\end{references}

%\end{document}

\section{Introduction}

In order to explain the main qualitative features of the Quantum Hall
effect one needs gauge invariance and impurities.  From the gauge 
invariance  one can obtain the quantization condition of the plateaux,
at least for the integer case\cite{Laughlin}\cite{Halperin}.  The
impurities are necessary to localize the states being filled 
on a plateau where $\sigma_{xy}$ remains constant.  

From the gauge arguments alone, it seems possible to infer that 
extended states exist at the center of impurity broadened Landau
bands\cite{Aoki}\cite{Halperin}\cite{Prange}.   
This is rather striking, since the issue of whether
states are localized or extended is normally a difficult problem
in Anderson localization, involving the study of Quantum Mechanics
in disordered potentials, at least in more than two dimensions. 
In two dimensions, with no magnetic field, states are in principle
always localized, no matter what the strength of the disorder\cite{Anderson}
\cite{Fisher}.  
It seems clear from these observations that there exists a competition
between localization due to impurity disorder  and the 
consequences of gauge invariance.
One expects then that the presence of the magnetic field can drive
impurity disorder irrelevant.

We recently presented a computation of the correlation length exponent
$\nu$ in a certain model with no disorder and obtained 
$\nu = 20/9$, in good agreement with experiments and numerical
simulations\cite{andre}.  
This exponent essentially followed from gauge invariance. 
In this paper we consider the effects of disorder in the above model.
We argue that to use conformal field theory techniques, one must
view the gauge potential as imaginary, whereas the usual scalar potential
is real. 
We present evidence that there may actually be two distinct 
universality classes depending on the relative strengths of the
gauge field verses impurity disorder.  In one universality class
all disorder is driven   irrelevant due to the disorder
in the gauge potential.  The gauge field is not normally thought of
as disordered, since the magnetic field is usually considered
uniform.  The disordered component of the gauge field can be
thought of as arising from the local electro-magnetic field due
to the random impurities as a source.    Alternatively,
conduction electrons find their way along
the most conductive paths, and the shapes of these paths can be
rather complicated in the presence of impurity disorder. 
Magnetic flux enclosed
by such paths is random, and this may perhaps effectively lead to 
disorder in the gauge field, as in the network 
model\cite{Chalk}. 

\section{Renormalization Group Analysis}

We consider fermions in two dimensions with hamiltonian $H$ and
second quantized action
\begin{equation}
\label{2.0}
S_{2+1} = \int dt d^2 x ~ \Psi^\dagger (i \d_t - H) \Psi 
\end{equation}
For the purposes of studying the consequences of disorder it is
convenient to Fourier transform in time 
\begin{equation}
\label{2.1}
\Psi (x,t) = \int d\vep ~ e^{i\vep t} ~ \Psi_\vep (x)  
\end{equation}
such that 
\begin{equation}
\label{2.1b}
S_{2+1} = \int d\vep \int d^2 x ~ \Psi^\dagger_\vep (\vep - H) \Psi_\vep 
\end{equation}
The functional integral is defined by $e^{iS_{2+1}}$.  For a fixed
$\vep$, one has a euclidean field theory with functional integral
defined by $e^{-S}$ where 
\begin{equation}
\label{2.2}
S = i \int d^2 x ~ \psi^\dagger_\vep (H-\vep ) \psi_\vep  
\end{equation}
With the hermiticity properties inherited from $2+1$
dimensions, one has that $S$ is anti-hermitian $S^\dagger = -S$.

\def\beq{\begin{equation}}
\def\eeq{\end{equation}}

We will study the hamiltonian 
\begin{equation}
\label{2.3} 
H = \inv{\sqrt{2}} (-i \d_x - A_x ) \sigma_x + \inv{\sqrt{2}} 
(-i\d_y - A_y ) \sigma_y + V(x,y) 
\eeq
where $A_\mu$ is the electro-magnetic vector potential and $V(x,y)$ 
is an impurity potential.  This kind of Dirac hamiltonian has been
considered before in the context of Quantum Hall transitions, with real
$V and A_\mu$, 
\cite{Ludwig}\cite{Chalker},  
however its meaning in \cite{andre} is rather different. 
Letting $\Psi = \left( \matrix{\psi_1 
\cr \psi_2 \cr } \right) $, and introducing the complex coordinates 
$z=(x+iy)/\sqrt{2}$, $\zbar = (x-iy)/\sqrt{2}$ and gauge fields
$A_z = (A_x -i A_y )/\sqrt{2}$, 
$A_\zbar = (A_x + i A_y )/\sqrt{2}$, the action is 
\begin{eqnarray}
\label{2.4}
S &=& \int \frac{d^2 x}{2\pi} 
\Bigr[ \psi_1^\dagger (\d_z -i A_z ) \psi_2 
+ \psi_2^\dagger (\d_\zbar - i A_\zbar ) \psi_1 
\\ \nonumber
&~& ~~~~~
-i V (\psi_1^\dagger \psi_1 + \psi_2^\dagger \psi_2 ) 
\Bigl] 
\end{eqnarray}
where we have dropped the $\vep$ term and suppressed the $\vep$ subscripts, 
since it was sufficient to study $\vep = 0$ in \cite{andre}.  
Using the hermiticity properties $\d_z^\dagger = - \d_\zbar$,
$A_z^\dagger = A_\zbar$ one verifies  $S^\dagger = -S$ when
$V$ is real.  

\def\beq{\begin{equation}}

We wish to study disorder in
 the above model using techniques from conformal field
theory.  When $V=0$, the model is conformally invariant and the action
is usually expressed in terms of left(L) and right(R) moving fermions.
Making the identifications:
\beq     
\label{2.5} 
\psi_1^\dagger = \psi_R^\dagger , ~~~~~
\psi_2 = \psi_R , ~~~~~
\psi_2^\dagger = \psi_L^\dagger, ~~~~~ 
\psi_1 = \psi_L 
\eeq
the action becomes the one appropriate to conformal field theory:
\begin{eqnarray}
\label{2.6}
S &=& \int \frac{d^2 x}{2\pi}
\BL  \psi_R^\dagger (\d_z -i A_z ) \psi_R 
+ \psi_L^\dagger (\d_\zbar - i A_\zbar ) \psi_L 
\\ \nonumber
&~& ~~~~~
-i V (\psi_R^\dagger \psi_L + \psi_L^\dagger \psi_R ) 
\BR 
\end{eqnarray}
The L, R designations come from the equations of motion when 
$V=0$:  $\psi_{L} = \psi_{L} (z)$, $\psi_R = \psi_R (\zbar)$.  

The conformal field theory is defined by functional integrals over 
$\psi_{L,R}, \psi_{L,R}^\dagger$, which is not evidently the same
as the functional integral over $\psi_{1,2}, \psi_{1,2}^\dagger$ since,
as can be seen from Eq. (\ref{2.5}), the hermiticity properties 
are not compatible.  This is most apparent in the conventional 
bosonization for the fermions, which reads:
\beg
\label{2.8}
\psi_L = e^{-i\phi_L}, ~~~~~ \psi_L^\dagger = e^{i\phi_L }, ~~~~~
\psi_R = e^{i\phi_R}, ~~~~~ \psi_R^\dagger = e^{-i\phi_R }
\end{equation}
where $\phi = \phi_L (z) + \phi_R (\zbar)$ is a free scalar field. 
The current coupled to the gauge field then has the following
bosonized expressions
\begin{eqnarray}
\label{2.9b}
j_z &=& \inv{2\pi} \psi_L^\dagger \psi_L = ~\frac{i}{2\pi} \d_z \phi
\\ \nonumber
j_\zbar &=& \inv{2\pi} \psi_R^\dagger \psi_R = - \frac{i}{2\pi} \d_\zbar \phi
\end{eqnarray}
which implies $j_\mu = \ep_{\mu\nu} \d_\nu \phi /2\pi$, where
$\ep_{12} = -\ep_{21} = 1$.

The bosonized action then takes the form 
\beg
\label{2.11b}
S = \int d^2 x ~ \[  \inv{8\pi} \d_\mu \phi \d_\mu \phi - \frac{i}{2\pi
}
\ep_{\mu\nu} \d_\nu \phi A_\mu \]
\end{equation}
The important point is that whereas
the kinetic term for the real scalar field is real, the term which
couples to the gauge field is imaginary if $A_\mu $ is real.  
Indeed, in \cite{andre}, it was shown that the Hall conductivity
$\sigma_{xy}$ computed from $S$ is real only if one makes an
analytic continuation which effectively makes $A_\mu$ imaginary
and thus renders $S$ real\footnote{In \cite{andre}, this corresponded
to letting the topological angle $\theta \to -i \theta$.}.

Based on the above considerations, we consider the effects of disorder
in the real potential $V$ and an imaginary vector potential $A_\mu$. 
The role of imaginary vector potentials in localization problems
has been previously recognized by Hatano and Nelson\cite{Nelson} 
and in \cite{Hatsugai}.  
Let $A_\mu \to A_\mu + i A_\mu^d$ where $A_\mu^d $ is a disordered
component of the gauge field, and $A_\mu = \d_\mu \chi$ is the background
value with the properties assumed in \cite{andre}.   The disordered
potentials are taken to have gaussian probability  distributions:
\begin{eqnarray}
\label{2.9}
P[A^d] &=& \exp \( -\inv{g_A} \int \frac{d^2x}{2\pi} ~ A_z^d A_\zbar^d 
\) 
\\ \nonumber
P[V]   &=& \exp \( -\inv{2g_V} \int \frac{d^2x}{2\pi} \( V(x) - V_0 \)^2 \)
\end{eqnarray}
where
$V_0$ is the mean value of $V(x)$ and $g_V, g_A$ represent {\it positive}
variances of the distribution.  

An effective action which incorporates the effects of the disorder
can be obtained using the supersymmetric method\cite{Efetov}.
  For our particular
problem this was carried out by Bernard\cite{Denis}.  Let us 
outline the main features.  Introducing
bosonic ghosts $\beta_{L,R}, \beta_{L,R}^\dagger$ that couple in the 
same way as the fermions, the functional integral over $V, A^d$ 
leads to the effective action:
\begin{equation}
\label{2.10}
S_{\rm eff} = S_{\rm cft} + \int \frac{d^2x}{2\pi} 
\( -i V_0 \CO_0 + g_V \, \CO_V + g_A \, \CO_A \) 
\eeq
where $S_{\rm cft}$ is the conformal field theory action for the fermions
plus a ghost action obtained from the fermionic one with the
replacement $\psi_{L,R} \to \beta_{L,R}$, and $\CO_0$ is
the operator the $V$ couples to in (\ref{2.6}) plus ghosts.   
The operators 
$\CO_{V,A}$ are:
\begin{eqnarray}
\label{2.11}
\CO_V &=& \inv{2} \( \psi_L^\dagger \psi_R + \psi_R^\dagger \psi_L 
+ \beta_L^\dagger \beta_R + \beta_R^\dagger \beta_L \)^2 
\\ \nonumber
\CO_A &=& \( \psi_R^\dagger \psi_R + \beta_R^\dagger \beta_R \) 
\( \psi_L^\dagger \psi_L + \beta_L^\dagger \beta_L \) 
\end{eqnarray}
The interactions can be written as a current-current perturbation of
the $OSP(2|2)$ super-current algebra\cite{Denis}.   

The operators $\CO_V, \CO_A$ do not form a closed operator algebra
and another term $g_M \, \CO_M$ in the lagrangian
  is generated under renormalization, with 
\beq
\label{2.12}
\CO_M = \inv{2} \( \psi_L^\dagger \psi_R - \psi_R^\dagger \psi_L 
+ \beta_L^\dagger \beta_R - \beta_R^\dagger \beta_L \)^2 
\eeq
Using the operator product expansions 
\begin{eqnarray}
\label{ope}
\psi_L^\dagger (z) \psi_L (0) &\sim& \psi_L(z) \psi_L^\dagger (0) \sim 1/z 
\\ \nonumber 
\beta_L (z) \beta^\dagger_L (0) &\sim& - \beta^\dagger_L (z) 
\beta_L (0) \sim 1/z
\end{eqnarray}
and similarly for $\beta_R , \psi_R$, one obtains the operator
product expansions
\begin{eqnarray}
\nonumber
\CO_V (z,\zbar) \CO_V (0) &\sim& \frac{-4}{z\zbar} \CO_V (0) 
\\ \nonumber 
\CO_M (z, \zbar ) \CO_M (0) &\sim& \frac{4}{z\zbar} \CO_M (0) 
\\ 
\CO_A (z,\zbar) \CO_M (0) &\sim&  
\frac{2}{z\zbar} \( \CO_V (0) + \CO_M (0) \) 
\\ \nonumber
\CO_A (z, \zbar) \CO_V (0) &\sim& 
\frac{2}{z\zbar} \( \CO_V (0) + \CO_M (0) \) 
\\ \nonumber
\CO_M (z,\zbar) \CO_V (0) &\sim& \inv{z\zbar} 
\( 4 \CO_A (0) -2 \CO_V (0) + 2 \CO_M (0) \)
\end{eqnarray}
From these  one can determine the lowest
order (1-loop) $\beta$eta functions, 
\beq
\label{beta}
\beta_g =  \frac{dg}{d\log R}  
\eeq
where
$R$ is a length scale\cite{Zamo}. 
One finds 
\begin{eqnarray} 
\beta_{g_V} &=& 4 g_V^2 + 4 g_M g_V - 4 g_A (g_M + g_V ) 
\\ \nonumber
\beta_{g_M} &=& -4 g_M^2 - 4 g_M g_V -4 g_A (g_M + g_V ) 
\\ \nonumber 
\beta_{g_A} &=&  -8 g_M g_V 
\end{eqnarray}
In \cite{Denis} the analogous result was obtained for a real vector
potential, which differs by signs $g_A \to - g_A$.  The special 
case $g_M + g_V = 0$ has some very interesting properties, namely
$\beta_{g_M + g_V} = 0$ when $g_M + g_V = 0$,
and the only non-zero $\beta$eta function
is $\beta_{g_A}$.  For the case of a real vector potential, one
can check numerically that renormalization group flows are drawn
to the line $g_M + g_V = 0$ in the UV.  
 This nearly conformal situation was studied in \cite{guru},
and is exactly soluble using $gl(N|N)$ supercurrent algebra.  

The change in sign for an imaginary vector potential has some important
consequences as far as the relevance or irrelevance of disorder.  
The $\CO_M$ operator which is generated corresponds to a term in
the hamiltonian $i M(x,y) \sigma_z $, which leads to a term in the
lagrangian $iM (\psi_1^\dagger \psi_1 - \psi_2^\dagger \psi_2 )$.  
This is anti-hermitian,  as it should be in the original 
$\psi_{1,2}$  formulation 
if $M$ is real.  Thus a positive $g_M$ corresponds
a variance in a gaussian distribution of $M$.  
Since the couplings $g$ represent variances of the potentials, we consider
the renormalization group flow when all couplings are initially positive.
The beta function $\beta_{g_M}$ then shows that $g_M$ is a marginally
irrelevant coupling, i.e. it decreases at larger distances.  At large
enough distance it is eventually driven to zero.  It will eventually
be driven negative using the 1-loop $\beta$eta functions, but since
this is unphysical   let us assume that 
 the flow is 
 cut-off when $g_M = 0$.    To get a qualitative
understanding of the renormalization group flow, 
  let us set  $g_M = 0$ since it
is marginally irrelevant.  One sees
that $g_A$ is then exactly marginal, i.e. it's $\beta$eta function vanishes.
What is interesting is that $g_V$ can be marginally relevant or irrelevant
depending on the initial strengths of $g_V$ verses $g_A$.  The two classes
are:

\bigskip

\noindent
{\bf Class A (Gauge Dominated)}  Here $g_A  >  g_V $.  
In this class $g_A$ drives $g_V$ to be
irrelevant.  Integrating the $\beta$eta function, one finds
\beq
\label{2.16}
g_V (R) = \frac{g_A}{ 1 + R^{4g_A} } 
\eeq
i.e. $g_V (R)$ decreases at large distances.  $g_V$ is more than
marginally irrelevant: it decreases much
faster than logarithmically because of the linear term in the
$\beta$eta function.      

\bigskip

\noindent
{\bf Class B (Impurity Dominated)}  Here $g_A <  g_V $, and
 $g_V$ is  relevant  
\beq
\label{2.17}
g_V (R) = \frac{g_A}{ 1 - R^{4g_A} } 
\eeq

\bigskip

To verify that this picture is not spoiled by a non-zero $g_M$, we
integrated the $\beta$eta functions numerically for an initial 
value of $g_M= 1/4$.   Indeed one sees two classes roughly 
separated by  $g_A =  g_V$.
In Class A all couplings decrease with increasing $R$, 
whereas in Class B $g_A, g_M$ decrease and $g_V$ increases.

\section{Discussion}

If  disorder is indeed irrelevant in Class A, there isn't much left for the
critical exponents of a transition to depend on.  What remains is
the constant mean value of the potential $V_0$, which is relevant, 
but clearly the exponent is independent of the strength of $V_0$.  
In \cite{andre} the impurity region was shrunk to a circular defect
of uniform strength and gauge arguments led to the exponent $\nu = 20/9$.
This is a reasonable proposal for the exponent governing Class A.  
For Class B the impurity disorder is  relevant and so it
is possible that one flows to a new fixed point with a different
exponent.  However the constant potential $V_0$ corresponds to a dimension 
$1$ operator and is more relevant.  If the exponent
in this class is indeed different, then the percolation type exponent of
$7/3$ is a possibility, since the impurities dominate and the
electrons must somehow percolate through islands of impurities\cite{seven}.   

There is a small amount of evidence for two universality classes in
both numerical simulations and experiments.  Huckestein's work gave
$\nu = 2.35 \pm .03$ which is consistent with $7/3$\cite{Huckestein}.  
On the other hand
Aoki-Ando's simulation gave $2.2 \pm .1$, closer to $20/9$\cite{Ando}\cite{Ando2}. 
It isn't clear yet whether there is any statistical significance to this. 
Furthermore Ando apparently doesn't incorporate randomness in the
gauge field directly.  
It would be very interesting to perform a simulation where one can tune
the strength of the gauge verses impurity disorder.

\section{Acknowledgments}

I would like to thank Denis Bernard, S. Guruswamy, B. Huckestein, 
 Andreas Ludwig and Jim Sethna for
discussions.  This work is supported in part by the National Young
Investigator Program of the NSF.

\bigskip

\vbox{
\vskip 1.5 truein
\epsfysize= 8cm
\epsfxsize= 8cm
\epsffile{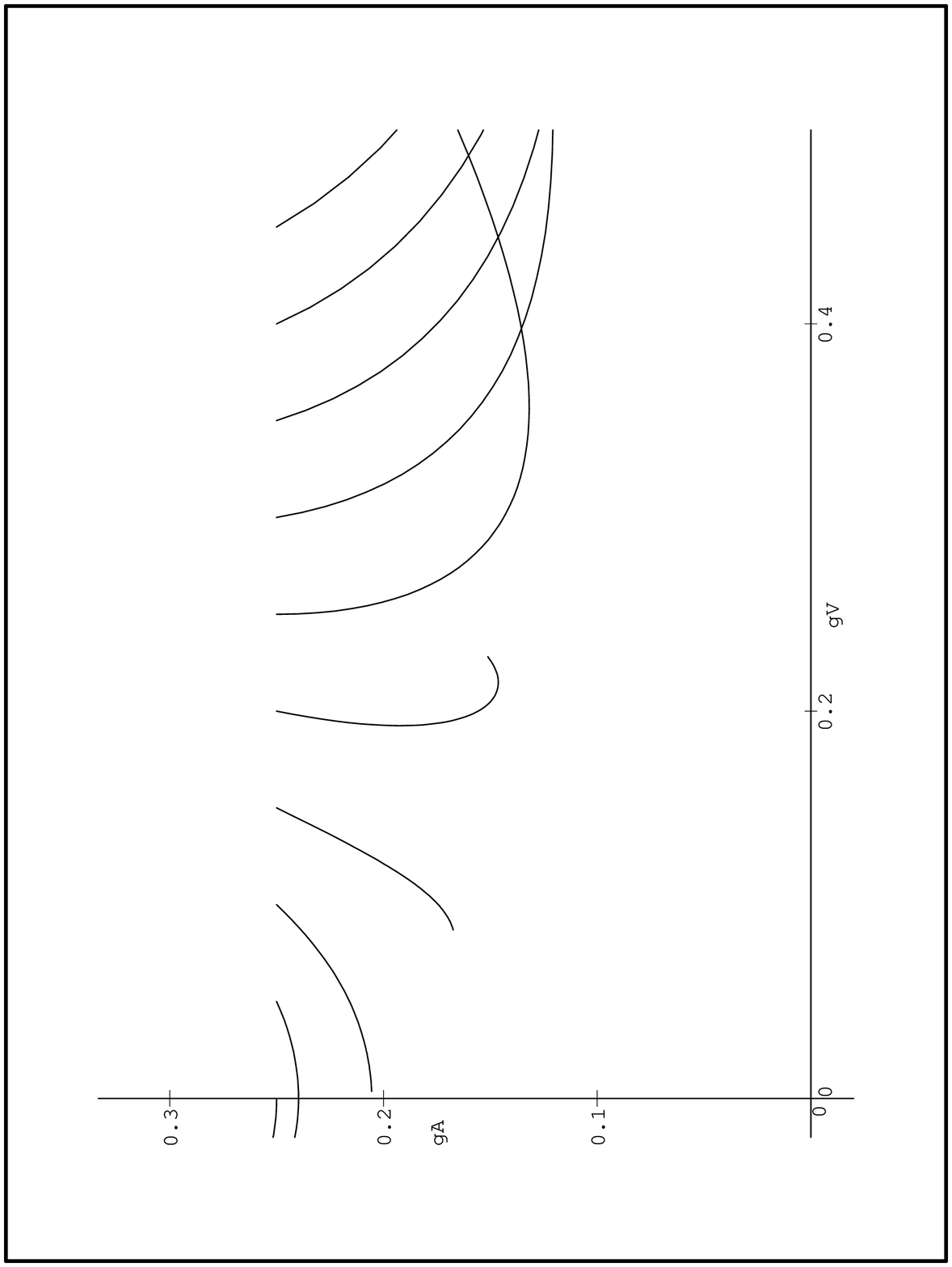}
\begin{figure}
\caption[]{\label{figure 1.}  Renormalization group flow with initial
values $g_A = g_M = .25$, for various values of $g_V$.  Increasing
length scale is from top to bottom.  
}
\end{figure}}

\end{document}